# Critical role of the sample preparation in experiments using piezoelectric actuators inducing uniaxial or biaxial strains

D. Butkovičová,[1] X. Marti,[2,1,3] [a)] V. Saidl,[1] E. Schmoranzerová-Rozkotová,[1] P. Wadley,[3,4] V. Holý,[1] and P. Němec[1]

[1] *Faculty of Mathematics and Physics, Charles University in Prague, Ke Karlovu 3, 121 16 Prague 2, Czech Republic*
[2] *Department of Materials Science and Engineering, University of California, Berkeley, California 94720, USA*
[3] *Institute of Physics ASCR v.v.i., Cukrovarnická 10, 162 53 Prague 6, Czech Republic*
[4] *School of Physics and Astronomy, University of Nottingham, Nottingham NG7 2RD, United Kingdom*

We report on a systematic study of the stress transferred from an electromechanical piezo-stack into GaAs wafers under a wide variety of experimental conditions. We show that the strains in the semiconductor lattice, which were monitored *in situ* by means of X-ray diffraction, are strongly dependent on both the wafer thickness and on the selection of the glue which is used to bond the wafer to the piezoelectric actuator. We have identified an optimal set of parameters that reproducibly transfers the largest distortions at room temperature. We have studied strains produced not only by the frequently used uniaxial piezostressors but also by the biaxial ones which replicate the routinely performed experiments using substrate-induced strains but with the advantage of a continuously tunable lattice distortion. The time evolution of the strain response and the sample tilting and/or bending are also analyzed and discussed.

## I. INTRODUCTION

Many recent advances in the design of novel materials commenced from the identification of competing energy scales among different degrees of freedom. The topology of the unit cell atoms (angles and distances), which is unique for each material, is ruled by the chemical bonds and determines the ground state for various physical properties (magnetic, electrical, optical, etc.). Lattice strains constitute a powerful tool to tune at wish the subtle equilibrium of any competing interactions and thus their corresponding ground state, even at apparently small strains of the order of $10^{-4}$. For instance, the ability to control the spin degree of freedom forms the basis for a construction of novel spintronic devices and – via the spin-orbit interaction – strain fields provide very sensitive means that can be used for this purpose.[1, 2]

A vast majority of experiments aiming at strain control are performed in epilayers that are strained due to the epitaxial growth on substrates with a slightly different lattice constant.[1] However, due to a limited and discrete choice of available substrates, this epitaxy-induced strains are ill suited to perform the desired *in situ* and continuous tuning of the strain. Alternatively, strains can be induced by a vice[2] but this technique has only a limited usage especially at cryogenic temperatures. On the other hand, piezoelectric materials are ideal for this purpose because they enable a voltage-control of the strain in a wide temperature range.[3] Despite the substantial amount of work and applications stemming from piezoelectric epitaxial layers, it is known that substrate clamping hinders a significant part of the stress which is not transferred into the functional layer. The use of single crystals or macroscopic piezo-actuator devices circumvents this issue and turned out to be a very useful in many material systems.[4-7] For example, in diluted magnetic semiconductors (DMS) – with

---

[a)] Electronic mail: xaviermarti@berkeley.edu

(Ga,Mn)As as a most well-known example – this method enabled the *in situ* control of magnetic anisotropy[7-9] and, consequently, the discovery of several new physical phenomena. In particular, the piezo-induced suppression of the temperature-related laser-pulse-induced precession of magnetization enabled us to observe the optical spin-transfer torque.[10] The piezo-electrically controlled magnetic anisotropy also enabled a demonstration of devices where up to 500% mobility variations for an electrical-current driven domain wall motions can be achieved.[11] However, the use of macroscopic actuators demands a reliable method to couple the actuator and the samples, which is an issue that we address in this paper.

For an interpretation of the results achieved with the piezo-stressors, it is of fundamental importance to characterize the strain generated along different directions in the crystal lattice. The strain measurements are routinely performed using the resistance strain gauges that are glued either to the piezo rod or to the investigated sample.[3, 7, 8] In this paper we report on a piezo-induced strain characterization by X-ray diffraction experiments where strains along *all three* directions – not only two as in the case of strain gauges – can be measured directly. We show that the piezo-induced strains in the lattice are strongly dependent on the wafer thickness and on the glue used to bond the sample to the piezoelectric actuator. Moreover, we revealed that for the uniaxial piezostressors, which were utilized in all previous experiments,[7-11] the lattice expansion along the piezo rod poling direction is accompanied by a lattice compression in the perpendicular directions – but *not only* in the sample plane but also (and considerably more) along the sample normal, which is a previously unrecognized effect.[3] In addition to this, we have measured the strains induced by the biaxial piezostressors that are producing in both in-plane directions strains with a similar magnitude (but with a opposite sign relative to that produced by the uniaxial piezostressors).

## II. EXPERIMENTAL

In our experiments, we used macroscopic lead zirconate titanate (PZT) piezotransducers. We investigated the strains induced both by the uniaxial piezo-stacks[12] (with a length of 9 mm and a $2 \times 3$ mm$^2$ cross section) and biaxial piezo-chips[13] (with a thickness of 2 mm and a $10 \times 10$ mm$^2$ cross section). As samples, we used commercial GaAs wafers with a size of about $2 \times 2$ mm$^2$ and an original thickness of 500 μm that were mechanically polished to a desired thickness (with a resulting thickness uncertainty of 10 μm). The polished wafers were bonded to piezostressors using 4 glues from different manufacturers (Epotek, Hysol, 3M, Norland).[14, 15] The piezostressors can be used to generate strains at various temperatures.[3] However, we limited our study to the room-temperature (RT) only due to several reasons. Firstly, we are aiming at construction of realistic spintronic devices where the RT operation is desirable. Secondly, glues usually do not transmit the whole stress produced by the piezo-actuator to the wafer – in Ref. 3, temperatures below 77 K were necessary for a full transmission of the strain. So, by the RT optimization we are in fact selecting the glue that provides good strain transmission from RT down to cryogenic temperatures. Finally, GaAs and PZT have a different thermal contraction and, consequently, cooling the sample to cryogenic temperatures induces sizable strains in GaAs even without bias applied to the piezostressors – in Ref. 7, cooling to 50 K produced in-plane biaxial tensile strain of $10^{-3}$ and a uniaxial strain of $\approx 10^{-4}$ at zero bias. So, by working at RT we reduce these bias-independent strains. The investigated piezo-actuators[12, 13] are poled components that enables a semi-bipolar operation mode – i.e., the voltage can be changed from -30 V to +150 V at RT. The produced strain scales monotonously with the bias, however, there is a pronounced hysteresis at 300 K.[3] To obtain a reproducible functionality of the piezo-stacks, we always performed an "initialization" procedure before the actual measurement (we repeated 3-times a bias sequence 0 V; +150; 0 V; -30 V).

The investigated wafers on piezostressors were installed on a sample holder of the Panalytical X'Pert Pro diffractometer. The set-up was armed with a Bartel's monochromator and an X-ray mirror on the incident side, and a crystal analyzer on the diffracted beam side. Two dimensional mapping of the scattering ($2\Theta$) and the rocking ($\omega$) angles were collected as a function of the piezo biasing voltage produced by the power source (PI, E-463). In these maps, $2\Theta$ direction probes (thru the Bragg's law) the lattice plane distance while the $\omega$ direction provides information on the sample quality (the diffraction peak broadening) and tilting (the diffraction peak location). As described in Ref. 16, several causes could be responsible for the peak broadening (for example the sample mosaicity, bending, defects, etc.). Moreover, we noticed sometimes a substantial ($\approx 0.1$ deg) tilt and (at least) doubling of the angular breath of the rocking curves upon bias-stressing the GaAs wafer back and forth. Consequently, by collecting two dimensional maps we circumvent the necessary re-alignment after applying the piezo voltage; we can extract all necessary information a posteriori by the software data analysis. In order to obtain information about out-of-plane and two perpendicular in-plane directions, we collected three reciprocal space maps per each experiment. The magnitude of the out-of-plane lattice parameter was obtained from the diffraction along the sample normal, (004), direction. The magnitudes of the in-plane lattice parameters were obtained from diffractions containing the in-plane components of the diffraction vector [either (113) and (1-13) or (115) and (1-15)].

### III. RESULTS AND DISCUSSION

In Fig. 1 we show the raw diffraction data measured along the (004) direction, which are directly related to the out-of-plane lattice constant of GaAs. In the upper section of the figure [parts (a) – (d)], which was measured for optimized sample preparation conditions (in the 200 μm thick wafer bonded by the Epotek glue), we show the desired functionality of the piezostressor: The diffraction peak is always rather narrow, its $2\Theta$-position changes reproducibly between -30 V [parts (a) and (c)] and +150 V [parts (b) and (d)], and the bias-induced sample tilt is rather small ($\Delta\omega < 0.02$ deg). For the contrast, we show in the lower section of Fig. 1 [parts (e) – (f)] what could be the effect of the piezostressor if the wafer thickness and/or the glue are not optimized: Even for –30 V the diffraction peak is considerably broader along the $\omega$ direction [cf. Fig. 1(a) and (e)], which is a signature of the glue-induced mosaicity and/or bending in GaAs. After application of +150 V, two diffraction peaks at different values of $2\Theta$ are clearly present [see Fig. 1(f)], which reveals that there are two parts of the wafer – one is strained by the piezostressor but the other one is not. Moreover, there is present also a rather large bias-induced sample tilt ($\Delta\omega \approx 0.08$ deg). Our data clearly highlights the importance of the wafer thickness and the glue selection – issues not addressed in the literature prior to this work. In Fig. 2(a) we show the wafer thickness dependence of the piezo-induced relative contraction of the lattice parameter (i.e., a compressive strain magnitude) in the out-of-plane direction for different glues.[14, 15] Clearly, the achieved strain depends significantly both on the wafer thickness and on the glue. From the tested combinations, the highest strain was obtained for the wafer thickness of 200 μm and the Epotek glue. Our further experiments also confirmed that this combination provides repeatedly a very similar strain as indicated by multiple data points in Fig. 2(a), which correspond to independently prepared samples on independent piezostressors. All the following results were obtained using this combination. In Fig. 2(b) we illustrate another feature of piezostressors that has to be considered in their utilization: the produced strain changes in time. In this figure we plot the measured $2\Theta$ scans as a function of the measurement time during the voltage switching. We started the experiment with a piezo stabilized at -30 V and measured two scans (each scan lasted $\approx 20$ min). Then we changed the bias voltage to +150 V and measured again two scans. After this we changed the voltage to -30 V etc. After the last change of the voltage we kept on measuring the data for 10 more hours. The conclusion from this experiment is that it takes $\approx 1$ hour for the strain to be fully stabilized in a wafer but after this time it is very stable. Moreover, these data also illustrate the stability of our X-ray diffraction measurement for long periods of time.

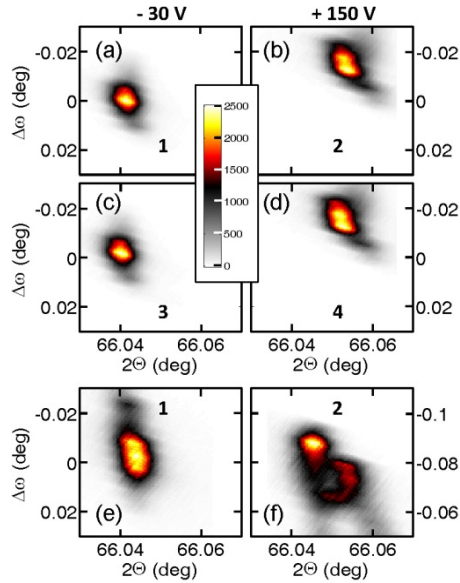

Fig. 1. Influence of a GaAs wafer thickness and a glue type on the *in situ* strain control by a uniaxial piezostressor. Two dimensional scans of scattering ($2\Theta$) and sample rocking ($\omega$) angles were measured for the (004) reflection at the optimized [200 um thickness and Epotek glue; parts (a) – (d)] and un-optimized [100 um thickness and 3 M glue; parts (e) – (f)] mounting combinations. Numbers indicate subsequent scattering experiments after biasing the sample with -30 V and +150 V. In all figures the same value of $\omega$ was subtracted for clarity; note that there is a different *y*-scale in part (f).

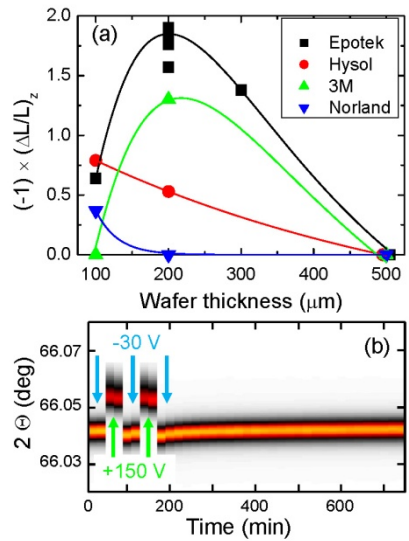

Fig. 2. (a) Dependence of the bias-induced relative contraction of the out-of-plane lattice parameter on the wafer thickness for different glues. The strains produced by the uniaxial piezostressor (points) were evaluated from the lattice parameters measured for the (004) reflections at -30V ($c^{-30V}$) and +150V ($c^{+150V}$) as $(\Delta L/L)_z = (c^{-30V} - c^{+150V})/c^{-30V}$. The lines are guides to the eye. (b) Time evolution of the piezo-induced strain after switching the applied bias, which is used to label the $2\Theta$ scans measured in the 200 um thick wafer bonded by Epotek glue to the uniaxial piezostressor.

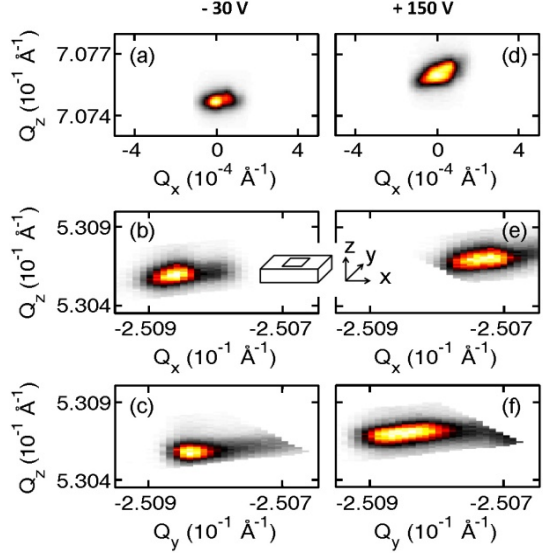

Fig. 3. Reciprocal space maps obtained at bias voltages -30 V (left column) and +150 V (right column) in the 200 um thick wafer bonded by Epotek glue to the uniaxial piezostressor. From top to bottom the data show reflections (004), (113) and (1-13); (113) corresponds to the poling direction of the stressor. Inset: Schematic depiction of the piezostressor with an attached wafer and of the coordinate system.

In Fig. 3 we show reciprocal space maps for GaAs wafer on an uniaxial piezostressor for applied biases of -30 V (left column) and +150 V (right column). The data shown in the first row were computed from the diffraction along the sample normal, (004), direction (see Fig. 1) and they reflect the bias-induced change of the out-of-plane lattice constant, which isproportional to $1/Q_z$ (the information about the in-plane parameters is not present in this diffraction direction and, therefore, $Q_x \approx 0$ in the corresponding graphs). The data in the second and the third rows depict the reciprocal space maps computed from the reflections (113) and (1-13) (i.e., along the piezo rod poling direction and perpendicular to it), respectively. From these reflections, the information about the out-of-plane [$y$-scale in Fig. 3(b)] and in-plane [$x$-scale in Fig. 3(b)] lattice parameters can be obtained. We note that we obtained the same bias-dependent change of the out-of-plane lattice parameter (strain) from all three reflections [(004), (113) and (1-13)], which confirms the credibility of our measurements. The bias-induced change of the in-plane lattice constant along the piezo poling direction (see the second row in Fig. 3) has a considerably larger magnitude and a opposite sign compared to that in the perpendicular in-plane direction (see the third row in Fig. 3). The bias-induced strains along different directions for this uniaxial piezostressor are summarized in Table I.

TABLE I. Bias-dependent lattice distortions in GaAs wafer with a thickness of 200 μm bonded by Epotek glue[14,15] to uniaxial[12] and biaxial[13] piezostressors. The relative distortions of the lattice along different directions $(\Delta L/L)_i$ ($i = x, y, z$) were computed from the lattice parameters measured by the X-ray diffraction experiment at piezo-bias of -30V ($a_i^{-30V}$) and +150V ($a_i^{+150V}$) as $(\Delta L/L)_i = (a_i^{-30V} - a_i^{+150V})/ a_i^{-30V}$.

| piezostressor | $(\Delta L/L)_x$ $(10^{-4})$ | $(\Delta L/L)_y$ $(10^{-4})$ | $(\Delta L/L)_z$ $(10^{-4})$ |
|---|---|---|---|
| uniaxial | +5.2 | -0.5 | -1.9 |
| biaxial | -5.0 | -5.0 | +3.1 |

In Fig. 4 we show the reciprocal space maps for the wafer on a biaxial piezostressor for applied biases of -30 V (left column) and +150 V (right column). By comparing Fig. 4 with Fig. 3, we can identify the differences between strains generated by the biaxial and uniaxial piezostressors (see also Table I). Firstly, the bias-induced change of the out-of-planelattice constant (strain) has an opposite sign in the biaxial piezostressor, which is connected with its different construction.[12, 13] Secondly, the strains measured in the two perpendicular in-plane directions have a same sign (which is opposite to that in the out-of-plane direction) and a very similar magnitude that confirms the biaxial character of this piezostressor. Finally, we investigated the consistency of the observed lattice distortions with that computed from the GaAs elastic constants[17] and lattice paramete1[8]. The in-plane and out-of-plane lattice parameters of the investigated GaAs wafer measured while biasing the piezostressor at +150 V (see Fig. 4) are $a_{meas}$ = 5.6517 Å and $c_{meas}$ = 5.6556 Å, respectively. For purely biaxial strains in a thin film grown on a substrate, the condition for the elastic equilibrium renders $c = a_{GaAs} - (2C_{12} / C_{11}) \times (a_{meas} - a_{GaAs}) = 5.6550(6)$ that is fully consistent with our direct measurements and, consequently, it verifies that the glue preserves the in-plane isotropic distortion of the piezostack.

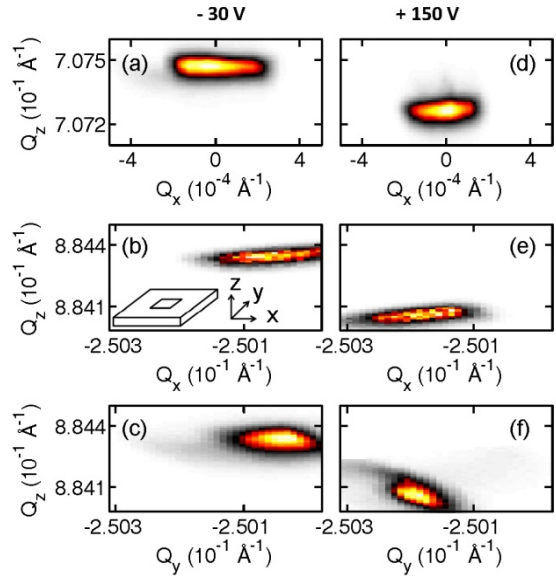

Fig. 4. Reciprocal space maps obtained at bias voltages -30 V (left column) and +150 V (right column) in the 200 um thick wafer bonded by Epotek glue to the biaxial piezostressor. From top to bottom the data show reflections (004), (115) and (1-15). Inset: Schematic depiction of the piezostressor with an attached wafer and of the coordinate system.

In conclusion, we have studied strains induced in GaAs wafer by piezostressors using X-ray diffraction experiment which enables a direct measurement of the bias-induced lattice constant change in all three crystallographic directions. We have tested several combinations of the wafer thicknesses and the glues and we have identified their combination that is reproducibly supplying the largest strain to the wafer at room temperature. Under these optimized conditions, we have measured strains produced not only by the frequently used uniaxial piezostressors but also by the biaxial ones which replicate the routinely performed experiments using substrate-induced strains but with the advantage of a continuously tunable lattice distortion. Our experiment thus broadens the spectrum of the experimentally available tools that can be used for the *in situ* tuning of the strain in semiconductors while signaling crucial and previously unaddressed experimental issues.

**ACKNOWLEDGMENTS**

This work was supported by the Grant Agency of the Czech Republic grants no. P204/12/0853 and P204/11/P339, and by the Grant Agency of Charles University in Prague grant no. 443011and SVV-2013-267306.

# Critical role of the sample preparation in experiments using piezoelectric actuators inducing uniaxial or biaxial strains: Supplementary material

D. Butkovičová,[1] X. Marti,[2,1,3] V. Saidl,[1] E. Schmoranzerová-Rozkotová,[1] P. Wadley,[3,4] V. Holý,[1] and P. Němec[1]

[1] *Faculty of Mathematics and Physics, Charles University in Prague, Ke Karlovu 3, 121 16 Prague 2, Czech Republic*
[2] *Department of Materials Science and Engineering, University of California, Berkeley, California 94720, USA*
[3] *Institute of Physics ASCR v.v.i., Cukrovarnická 10, 162 53 Prague 6, Czech Republic*
[4] *School of Physics and Astronomy, University of Nottingham, Nottingham NG7 2RD, United Kingdom*

In this supplementary material we provide further information about the glues that were used for bonding the GaAs wafers to the piezostressors. We used 3 different two component epoxy glues (Epotek, Hysol, 3M) and 1 one component glue (Norland) that are specified in Table SI. We selected these glues because they have a very large shear strength and hardness as apparent from Table SII where their selected parameters as listed. All glues were temperature cured at 120°C for 15 minutes. After the curing, the glue layer thicknesses were between 20 um and 100 um for all the glues.

**TABLE SI**. List of glues used for bonding the GaAs wafers to the piezostressors.

| producer | product name | abbreviated name used in this paper |
|---|---|---|
| Epoxy Technology | Epotek H70E | **Epotek** |
| Loctite | Hysol 9492 | **Hysol** |
| 3M | Scotch-Weld 9323B/A | **3M** |
| Norland Products | NEA 121 | **Norland** |

**TABLE SII**. Selected parameters of the glues as specified by the manufacturers.

| glue | **Epotek** | **Hysol** | **3M** | **Norland** |
|---|---|---|---|---|
| shore D hardness | 83 | 80 | | 85 |
| shear strength at 23°C (psi) | > 2 000 | 4 500 | 5 250 | 3 500 |
| operating temperature (K) | 218 - 573 | 233 - 353 | 218 - 355 | 123 - 423 |
| thermal expansion ($10^{-5}$ K$^{-1}$) | 1.5 | 6.3 | | |
| thermal conductivity (W/mK) | 0.9 | 0.3 | | |